\documentclass[%
 aip,
 amsmath,amssymb,
 reprint,%
]{revtex4-1}

\usepackage{graphicx}
\usepackage{dcolumn}
\usepackage{bm}

\usepackage[utf8]{inputenc}
\usepackage[T1]{fontenc}
\usepackage{mathptmx}
\usepackage{etoolbox}
\usepackage{subfigure}
\usepackage{hyperref}
\usepackage{dcolumn}
\usepackage{amsmath}

\makeatletter
\def\@email#1#2{%
 \endgroup
 \patchcmd{\titleblock@produce}
  {\frontmatter@RRAPformat}
  {\frontmatter@RRAPformat{\produce@RRAP{*#1\href{mailto:#2}{#2}}}\frontmatter@RRAPformat}
  {}{}
}%
\makeatother
\begin{document}

\preprint{AIP/123-QED}

\title[Sample title]{Mobile chemical cage: Revealing the origin of anomalous lithium diffusion in liquid $Li_{17}Pb_{83}$ alloy}
\author{Sensen Lin}
\affiliation{%
	College of nuclear science and technology, Harbin Engineering University, 150001 Harbin, China.
}%
 
\author{Yang Gao}
\altaffiliation{Electronic mail: gaoyang@hrbeu.edu.cn}
\affiliation{%
	College of nuclear science and technology, Harbin Engineering University, 150001 Harbin, China.
}%

\author{Yongheng Lu}%
\affiliation{ 
Metallurgical Research Institute, China North Nuclear Fuel Co.,Ltd, 014000 Baotou, China
}%

\author{Yongkuan Zhang}
\altaffiliation{Electronic mail: ykzhang@hrbeu.edu.cn}
\affiliation{%
	College of nuclear science and technology, Harbin Engineering University, 150001 Harbin, China.
}%

\author{Yiqiang Sun}
\affiliation{ 
	Metallurgical Research Institute, China North Nuclear Fuel Co.,Ltd, 014000 Baotou, China
}%

\date{\today}

\begin{abstract}
The high-temperature performance of liquid $Li_{17}Pb_{83}$, a key fusion reactor material, is governed by its atomic-scale dynamics. Using ab initio molecular dynamics, we discover that lithium diffusion is not free but confined within cages formed by lead atoms, a phenomenon we term the chemical cage effect. Structurally, RDF and CSRO analyses confirm a stable local environment where Li is preferentially surrounded by Pb. Dynamically, the MSD and NGP reveal anomalous, heterogeneous lithium diffusion characterized by repeated cage-breaking events. The double-exponential relaxation of the Li-Pb bond probability further distinguishes the escape dynamics of Li from surface and bulk cages. ELF and DOS analyses identify the polar covalent Li-Pb bond as the electronic origin of this cage. This study establishes the chemical bond-directed synergistic cage effect as the core mechanism in $Li_{17}Pb_{83}$, moving beyond traditional geometric constraint models and providing a new paradigm to understand transport in multi-component liquid alloys. 
\end{abstract}

\maketitle

\section{Introduction}
Fusion energy \cite{fusionenergy1,fusionenergy2,fusionenergy3} stands as a pivotal frontier in the global pursuit of sustainable, low-carbon power, with flagship projects like the International Thermonuclear Experimental Reactor (ITER) \cite{ITER1,ITER2} leading the charge toward practical fusion energy deployment. Within such reactors, liquid lithium-lead alloys—particularly the prototypical $Li_{17}Pb_{83}$ alloy \cite{2008Lead,2016Volumetric,R2017Volumetric}. have emerged as irreplaceable multitasking materials: they simultaneously function as tritium breeders (critical for sustaining the fusion fuel cycle) \cite{TBM1,TBM2} and high-temperature coolants (ensuring reactor component stability under extreme conditions). However, the atomic-scale dynamics that directly govern the performance of $Li_{17}Pb_{83}$ under reactor-relevant high temperatures remain largely uncharted. A key unresolved question is the diffusion mechanism of lithium (Li) atoms: as the primary contributor to tritium transport and thermal conductivity, Li’s anomalous diffusion behavior in the Pb-dominated matrix has long lacked a clear microscopic explanation, creating a bottleneck for optimizing $Li_{17}Pb_{83}$-based blanket materials.

In the study of dense liquid dynamics, the cage effect which is a core concept of Mode Coupling Theory (MCT) \cite{2013Nonequilibrium,2015Unraveling,2018Mode} become the dominant framework for interpreting atomic localization and anomalous diffusion. First proposed to describe supercooled liquids and colloids \cite{2015Shear,2016Cage}, the MCT derived cage effect posits that an atom is temporarily confined by the collective spatial arrangement of its neighbors, giving rise to non-Gaussian diffusion behavior extensively validated via theory, simulation and scattering experiments \cite{2015Liquid,0Catching,2020Experimental,2024Early,Juan2004Kinetics,2015Two}. A defining feature of traditional MCT is its reliance on geometric constraints and volume packing effects. The cage is conceptualized as a transient trap arising from excluded volume by hard-sphere-like spatial repulsion between atoms \cite{2006Volume,2016Granular}. Yet, despite its success in simple systems, traditional MCT faces limitations when applied to high-temperature, multi-component liquid alloys like $Li_{17}Pb_{83}$. It explicitly neglects chemical bond interactions and treats atoms as neutral hard spheres, failing to account for the strong Li-Pb chemical affinity and electronic effects that define alloy behavior. Recent extensions of MCT to metallic liquids \cite{Wong_2015,MCT} have only reinforced this gap while they observe density-dependent cage dynamics and hidden dynamic transitions in hard-sphere-like metallic systems, they still overlook the role of chemical bonding in cage formation, leaving the unique behavior of $Li_{17}Pb_{83}$ unaddressed.
Compounding this theoretical limitation, existing research on $Li_{17}Pb_{83}$ itself has focused primarily on macroscale thermophysical properties rather than atomic-scale dynamic mechanisms. Early studies by I.E. Lyublinski et al. \cite{1995Numerical} quantified transition metal solubility in $Li_{17}Pb_{83}$ using coordination cluster theory, while Xianglai Gan et al. \cite{GAN20142946} later characterized its structural and thermodynamic properties via molecular dynamics. S.G. Khambholja and A. Abhishek \cite{2020Structure} further reported temperature-dependent structural factors for Pb-Li eutectics, noting deviations from hard-sphere behavior due to mass disparities between Pb and Li. Most recently, Wenyi Ding et al. \cite{DING2026117011} investigated alumina coating corrosion in liquid LiPb, focusing on interface interactions rather than bulk diffusion. Collectively, these works confirm $Li_{17}Pb_{83}$’s structural uniqueness but stop short of linking its chemical bonding to dynamic behavior. None have identified the origin of Li’s anomalous diffusion, nor have they addressed the limitations of MCT in describing this alloy’s cage effect.

To fill these gaps, we present a systematic study of liquid $Li_{17}Pb_{83}$ at 800 K using AIMD simulations. Our findings challenge the traditional MCT paradigm, Li diffusion is not constrained by geometric volume packing, but by a chemically driven cage effect where Li atoms are dynamically surrounded by Pb atoms via polar covalent Li-Pb bonding. We further uncover a previously unreported cage coupling phenomenon via four-point dynamic structure factor analysis. It was shown that cage-breaking and cage-reforming events are not isolated but spatially correlated across the Pb matrix. By identifying the electronic origin of the chemical cage and quantifying cage coupling, this work extends MCT to account for chemical specificity in high-temperature liquid alloys. Ultimately, our results provide a new atomic-scale framework for understanding transport in multi-component liquid alloys, with direct implications for optimizing fusion reactor blanket materials.

\section{Theory and calculation}
Ab initio molecular dynamics (AIMD) simulations were performed using the CP2K package, which employs the Gaussian and Plane Waves (GPW) method \cite{1997A,GPW}. This approach was selected as it is uniquely suited for accurately modeling the complex, multi-component liquid $Li_{17}Pb_{83}$ system. 
The GPW method, as implemented in the CP2K package \cite{CP2K2005,CP2K2014,CP2K2020,CP2K2022}, was employed for all AIMD simulations due to its superior efficiency for large, periodic systems containing hundreds of atoms. The GPW method overcomes the limitations of pure Gaussian orbital and pure plane-wave bases by using Gaussian-type orbitals (GTOs) for the representation of the Kohn-Sham wavefunctions and an auxiliary plane-wave basis for the electron density. This dual-basis approach allows CP2K to efficiently handle both the localized nature of covalent bonds and the long-range Coulomb interactions in periodic systems. Within the GPW formalism, the DFT total energy functional is minimized with respect to the Kohn-Sham orbitals $\left\{ {{\psi _i}} \right\}$ under orthogonality constraints:
\begin{equation}
{E_{tot}} = \min \left\{ {{E_{KS}}\left[ {\left\{ {{\psi _i}} \right\}} \right] + {E_{OV}}\left[ {\left\{ {{\psi _i}} \right\}} \right]} \right\}
\end{equation}
The Kohn-Sham energy ${{E_{KS}}}$ is evaluated as:
\begin{equation}
	\begin{split}
		E_{\text{KS}} &= \underbrace{\sum_i \left\langle \psi_i \left| -\frac{1}{2}\nabla^2 \right| \psi_i \right\rangle}_{T_s} + \underbrace{\int V_{\text{ext}}(r) n(r) dr}_{E_{\text{ext}}} + \\
		& \quad \underbrace{E_H[n]}_{\text{Hartree}} + \underbrace{E_{\text{XC}}[n]}_{\text{XC}} + \underbrace{E_{\text{PP}}}_{\text{Pseudopotential}}
	\end{split}
\end{equation}
Here, the electron density $n(r)$ is expanded in the plane-wave basis, enabling the use of Fast Fourier Transforms (FFT) to compute the Hartree energy ${{E_H}\left[ n \right]}$ with $O\left( {N\log N} \right)$ scaling. This central feature, along with the avoidance of explicit four-center integrals, reduces the overall computational complexity to nearly $O\left( {{N^2}} \right)$, making CP2K exceptionally suited for our study of the liquid $Li_{17}Pb_{83}$ system. To accurately describe the strong electronic correlations in the Pb-6s orbitals, the DFT+U approach following Dudarev's formulation was applied:
\begin{equation}
{E_{DFT + U}} = {E_{DFT}} + \frac{{{U_{eff}}}}{2}\sum\limits_\sigma  {Tr} \left[ {{n^\sigma }\left( {1 - {n^\sigma }} \right)} \right]
\end{equation}
where ${{n^\sigma }}$ is the occupation matrix for the Pb-6s states and ${U_{eff}}$ is the effective on-site Coulomb parameter. Interatomic forces were computed via the Hellmann-Feynman theorem to ensure accurate AIMD trajectories.

A simulation cell of $Li_{17}Pb_{83}$ 100 atoms (17 Li and 83 Pb) was constructed through Avogadro \cite{2012Avogadro} with periodic boundary conditions applied in all directions. The initial triclinic cell was a=22 Å, b=22 Å, c=22 Å, angles $\alpha$=90$^ \circ $, $\beta$=90$^ \circ $, $\gamma$=90$^ \circ $, and a volume of 10648 Å$^3$. After optimizing by CP2K, had dimensions of a=21.91500 Å, b=21.91600 Å, c=21.91300 Å, angles $\alpha$=90$^ \circ $, $\beta$=90$^ \circ $, $\gamma$=90$^ \circ $, and a volume of 10524.58 Å$^3$. Input file with NPT ensemble was prepared using Multiwfn \cite{2012Multiwfn,2024A}, and structural visualized with VESTA \cite{VESTA}.
Convergence tests were rigorously conducted. The plane-wave cutoff energy was tested from 100 to 600 Ry. 400 Ry was chosen as it yielded the lowest total energy (-417.23950873836793 Ha) and met the convergence criterion of $\Delta E < 0.1$ meV. With the cutoff fixed at 400 Ry, k-point convergence was tested using $1 \times 1 \times 1$, $2 \times 2 \times 2$, $3 \times 3 \times 3$ meshes. The $1 \times 1 \times 1$/$\Gamma $ point was sufficient, achieving energy convergence better than 1 eV/atom. The self-consistent field (SCF) and geometric optimization force convergence thresholds were set to $1 \times {10^{ - 6}}$ Ha and $4.5 \times {10^{ - 4}}$ Ha/Bohr, respectively. 
Geometry relaxation was first performed using CP2K and Multiwfn \cite{2012Multiwfn,2024A} to optimize atomic coordinates and cell parameters. Subsequently, AIMD simulations in the NPT ensemble were conducted at 800 K with CSVR thermostat, as implemented in CP2K 2025.1. The simulation spanned 10 ps with a 1.0 fs time step. To enhance efficiency, the ASPC extrapolation method and an SCF convergence threshold of $1 \times {10^{ - 5}}$ Ha were used. Atomic coordinates, velocities, and forces were recorded every 10 steps for post-processing. 
To validate the reliability of the calculation, experimental/theoretical benchmarking was performed. From the NPT simulation trajectory, the average cell volume $\langle V \rangle$ was computed by averaging the equilibrium volume (after discarding the initial 1 ps equilibration period to exclude transient effects). The system density $\rho$ was calculated to compare with literature data for liquid $L{i_{17}}P{b_{83}}$ which Xianglai Gan et al. reported a density of 9.8 $g/c{m^3}$ via molecular dynamics simulations, while S.G. Khambholja and A. Abhishek noted consistency between theoretical and experimental densities (within 2\% error) for Pb-Li eutectics. Our calculated density (10.78 $g/c{m^3}$) falls within this range, confirming the accuracy of the simulation cell, force field parameters, and NPT ensemble setup to validate the modeled system correctly reproduces the macroscale thermophysical properties of liquid $L{i_{17}}P{b_{83}}$ before subsequent structural and dynamic analyses. The structural and dynamic analyzes were performed on the AIMD trajectories. The Electron Localization Function (ELF) \cite{ELF2005,ELF2016} was calculated with a 600 Ry grid cutoff to analyze chemical bonding.

\section{Results and discussion}

\subsection{Structural evidence of chemical cage}
\begin{figure}[h]
	\centering
	\subfigure[]{
		\includegraphics[width=1.5in]{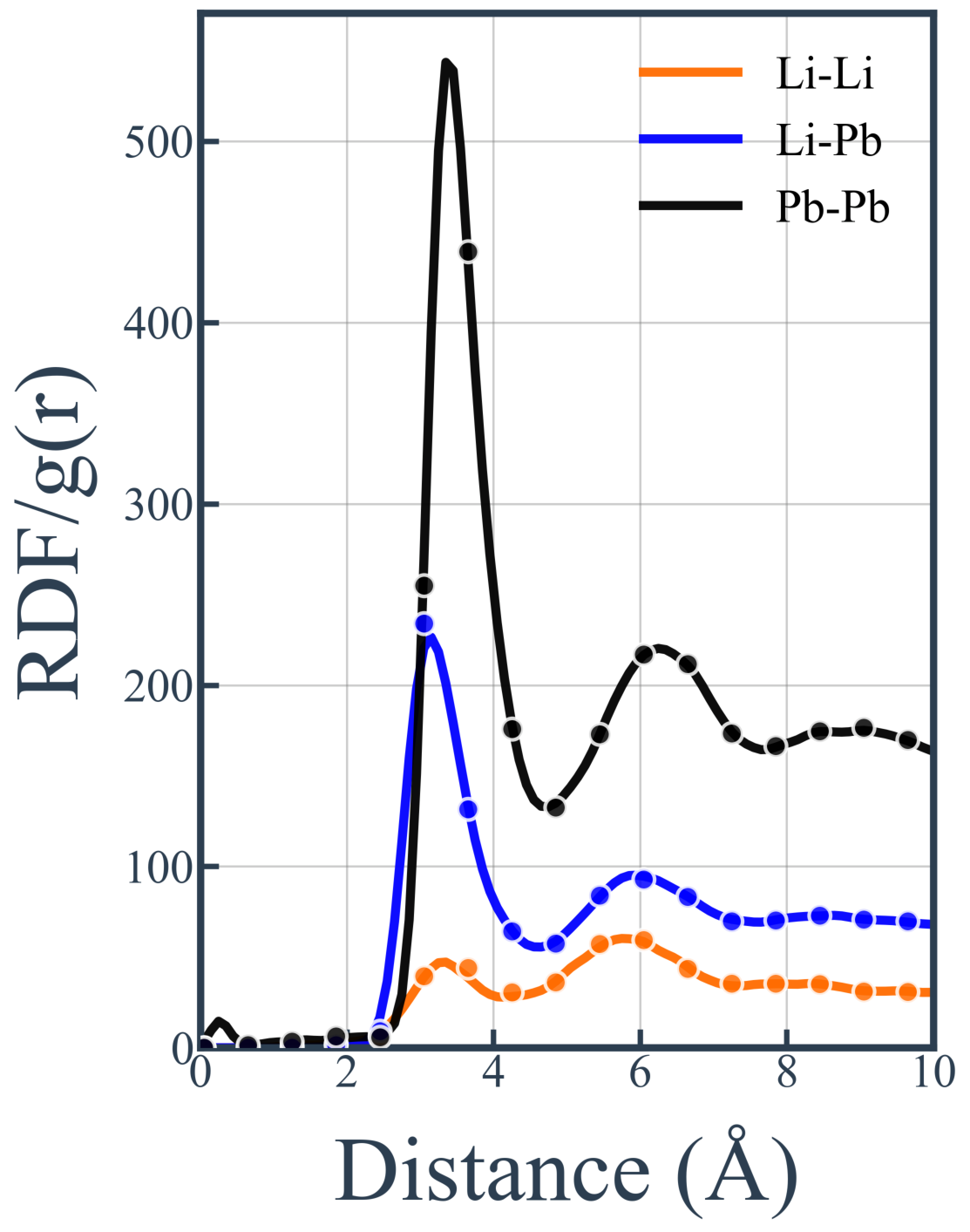}
		\label{pdf_li17pb83_white_bg}
	}
	\quad
	\subfigure[]{
		\includegraphics[width=1.5in]{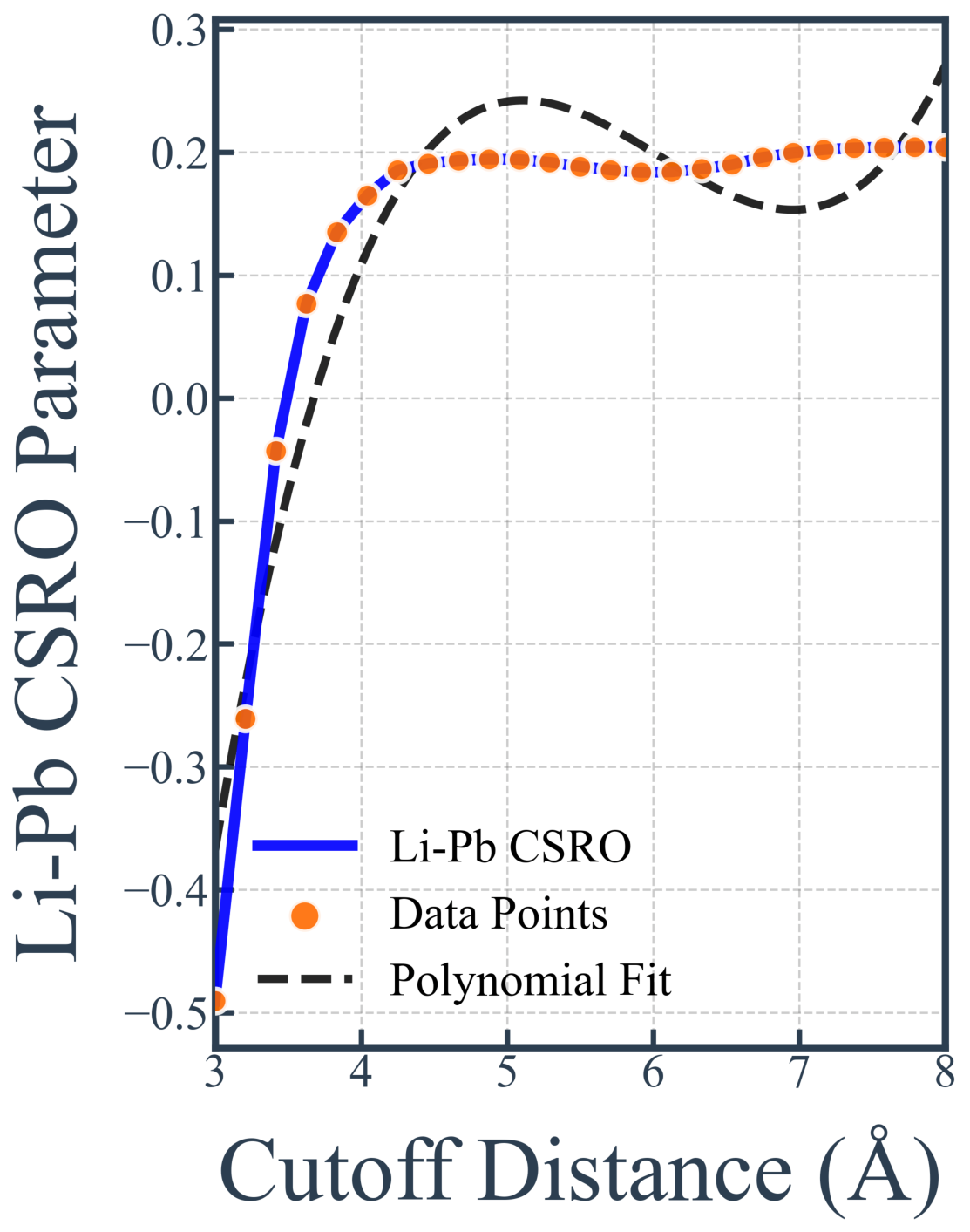}
		\label{csro_li17pb83_white_bg}
	}
	\caption{(a) Radial distribution function(RDF) of different pairs (b) Li-Pb CSRO parameter.}\label{rdf}
\end{figure}

To structurally confirm the existence of the proposed chemical cage, the radial distribution function (RDF) of the liquid $L{i_{17}}P{b_{83}}$ alloy in Figure. \ref{pdf_li17pb83_white_bg} provides direct structural evidence for the formation of a local cage around Li atoms. The dominant first peak for Pb-Pb pairs at 3.5 Å with a significantly higher intensity than those of Li-Li and Li-Pb pairs, indicates that Pb atoms tend to interconnect to form a persistent and well-defined network. This Pb-rich network constitutes the structural framework of the cage.
Conversely, the low and broad first peak of Li-Li pairs suggests that Li atoms are largely isolated from each other and dispersed within the Pb matrix. The larger half-width of the Li-Li peak signifies a more discrete distribution and greater positional fluctuation of Li atoms, which is a structural manifestation of their active thermal motion within the confinement of the Pb cages.
At distances beyond 5 Å, all RDFs converge towards unity, confirming the long-range disorder characteristic of a liquid. However, the more pronounced deviation of the Pb-Pb RDF at longer distances highlights the extended influence of its network structure, which defines the spatial extent of the local cages governing Li diffusion. The evolution of the Li-Pb chemical short-range order (CSRO) parameter with cutoff distance shown in Fig. \ref{csro_li17pb83_white_bg}, serves as a quantitative fingerprint of the local cage structure. The parameter's rapid rise and stabilization at a positive value within the critical 4-8 Å range is particularly revealing.
This stable plateau signifies that the preferred "Li surrounded by Pb" coordination environment is not a fleeting first-shell phenomenon but a robust and extended local structure. The 4 Å mark effectively defines the inner boundary of the formed cage, where the primary coordination shell is fully encompassed. The persistence of this order out to 8 Å demonstrates that the cage's influence extends across multiple atomic shells, giving it a well-defined spatial dimension.
The positive value of the CSRO parameter is crucial, as it quantitatively confirms the preferential segregation of Pb atoms around Li, forming the chemical basis of the cage. This stable chemical ordering presents a potential energy landscape that Li atoms must navigate. Therefore, CSRO profile does not merely indicate a static structure, it defines the thermodynamic cage that confines the Li atoms. The subsequent diffusion of Li characterized by anomalous and heterogeneous dynamics can be fundamentally understood as the process of breaking and reforming this specific CSRO-defined local environment.

\begin{figure}[h]
	\centering
	\subfigure[]{
		\includegraphics[width=1.5in]{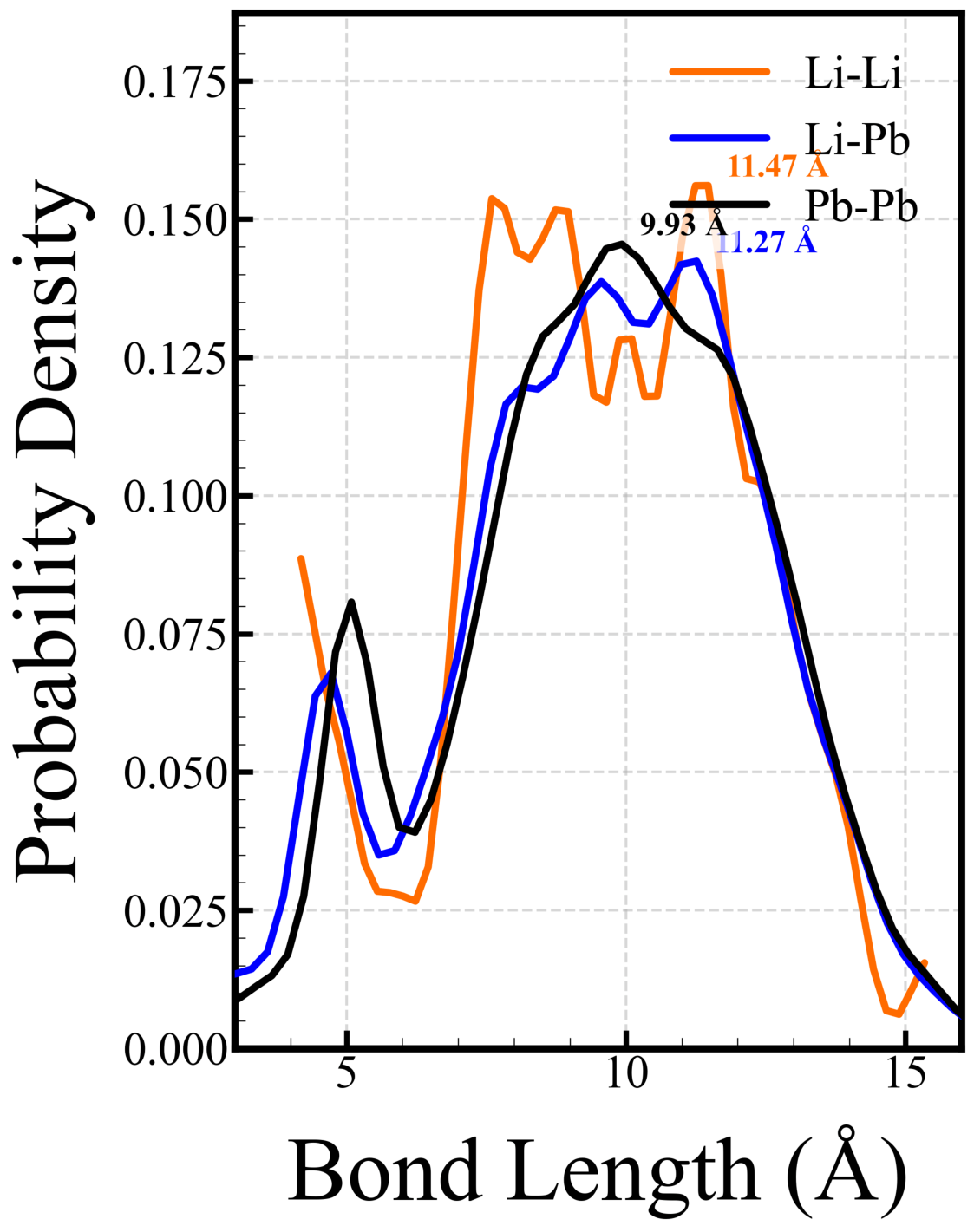}
		\label{all_pairs_bond_distribution_SCI}
	}
	\quad
	\subfigure[]{
		\includegraphics[width=1.5in]{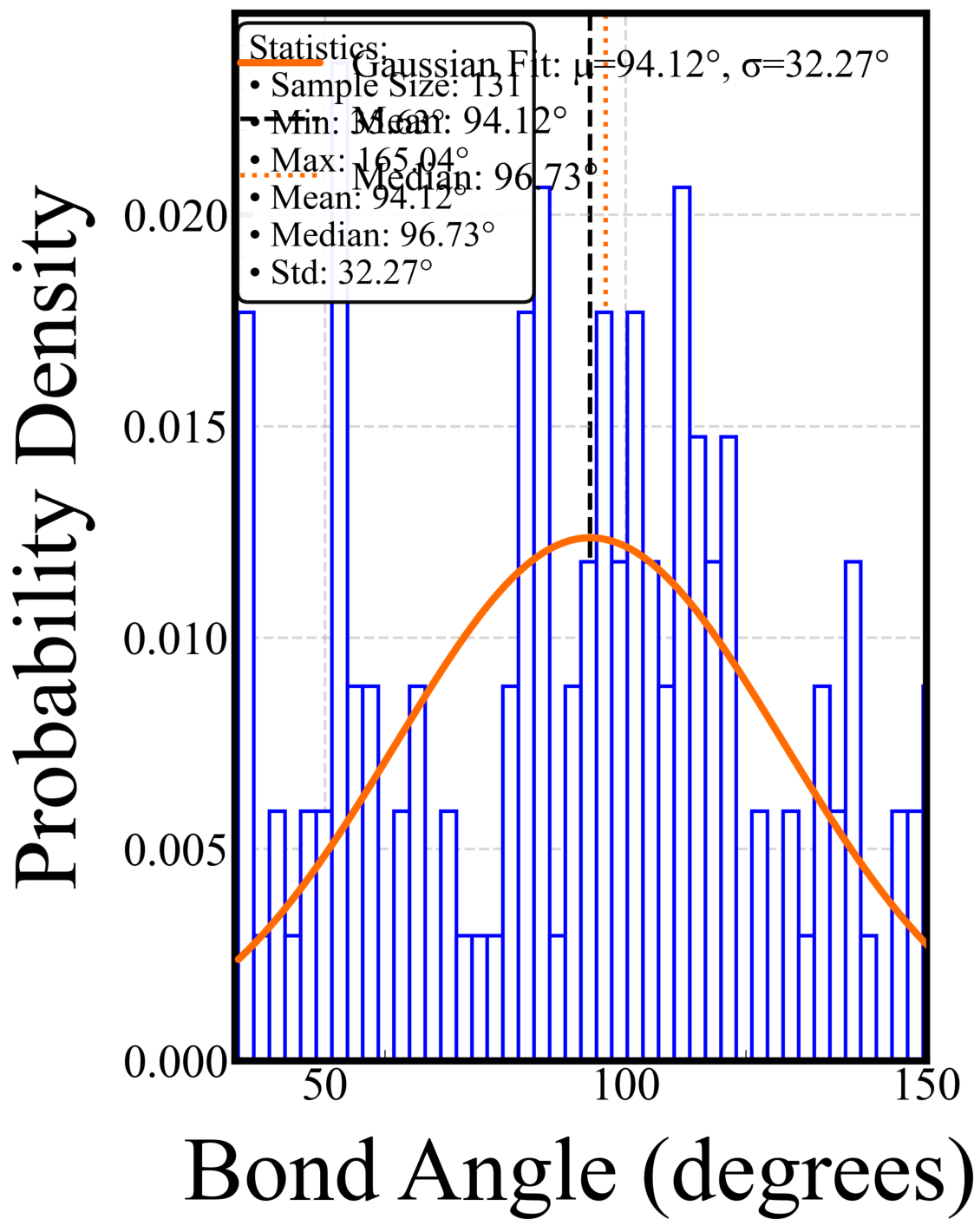}
		\label{angle_distribution_SCI.eps}
	}
		\quad
	\subfigure[]{
		\includegraphics[width=3.1in]{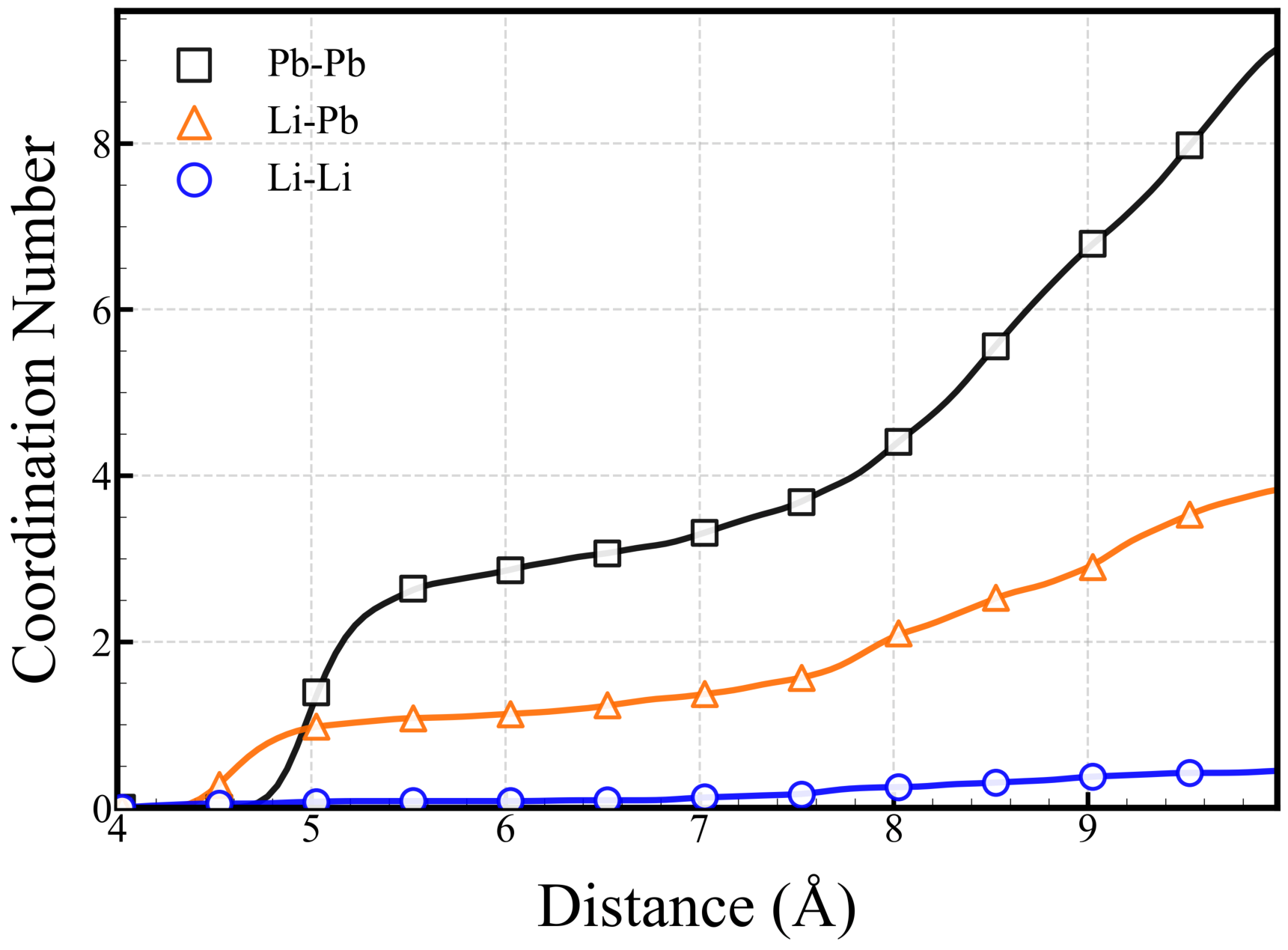}
		\label{coordination_curves_SCI}
	}
	\caption{(a) Bond length distribution (b) Bond angle distribution (c) Coordination number.}\label{bond}
\end{figure}

The bond length and bond angle distributions provide a definitive geometric characterization of the soft, dynamic cage confining the lithium atoms.
The bond length distribution in Fig. \ref{all_pairs_bond_distribution_SCI} reveals the most probable distances for different atomic pairs, which are fundamental to the cage's dimensions. The Li-Pb bond length peaks at 9.14 Å, defining the characteristic radial distance from a central Li atom to its surrounding Pb cage wall. The broad distribution of Li-Pb bond lengths signifies that this distance is not fixed but exhibits significant fluctuations, a hallmark of the cage's flexibility and anharmonicity. In contrast, the sharper Pb-Pb bond length distribution (peak at 9.41 Å) reflects a more well-defined and stable connectivity within the Pb network that forms the cage's scaffolding. The presence of a distinct but broadened Li-Li peak confirms that Li atoms are predominantly isolated from one another, as they are individually encapsulated within their Pb-dominated coordination environments.
A non-rigid and fluctuating cage is unequivocally confirmed by the Li-Pb-Li bond angle distribution in Fig. \ref{angle_distribution_SCI}. The average angle of 93.73$^ \circ $ strongly deviates from the ideal tetrahedral (109.5$^ \circ $) or octahedral (90$^ \circ $) coordination, immediately revealing a highly distorted and disordered local geometry. Crucially, the exceptionally large standard deviation of 33.71$^ \circ $ is a direct metric of the cage's extreme dynamic disorder. The angles are not locked but undergo continuous, large-amplitude rearrangements, meaning the cage perpetually breathes and reshapes.
In synthesis, these two structural descriptors paint a coherent and vivid picture. The cage is a soft and polydisperse coordination structure with a characteristic but variable size (bond lengths) and a highly irregular, fluctuating shape (bond angles). This inherent structural flexibility is not a minor detail but is fundamental to the diffusion mechanism. It is the stochastic, large-scale opening of these distorted angles, coupled with bond length fluctuations, that creates transient pathways for a Li atom to break its coordination bonds and initiate an escape from its confinement, directly leading to the anomalous diffusion behavior observed in the dynamics analysis. The coordination number (CN) profiles and the CSRO parameter jointly delineate the spatial and chemical structure of the local cage.
As shown in Fig. \ref{coordination_curves_SCI}, the CN of all atomic pairs increases with distance and eventually plateaus, consistent with the short-range order characteristic of liquids. Crucially, the Pb-Pb coordination curve grows the fastest and reaches the highest final value 18, which quantifies the dense, interconnected network of Pb atoms that forms the rigid scaffold of the cage. In contrast, the Li-Li CN is the lowest, confirming that Li atoms are effectively isolated from each other and dispersed within the Pb-dominated matrix. The intermediate growth rate and final value of the Li-Pb CN directly reflect the dominant local bonding environment, each Li atom is preferentially coordinated by multiple Pb atoms.

\subsection{Dynamics of cage breath and coupling behavior}
\begin{figure}[h]
	\centering
	\includegraphics[width=3.3in]{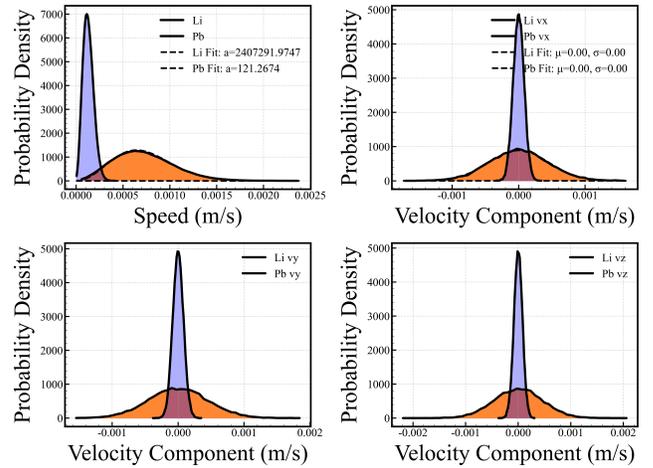}
	\caption{(a) Speed distribution (b) Velocity component distribution (x-direction) (c) Velocity component distribution (y-direction) (d)Velocity component distribution (z-direction).}
	\label{velocity_analysis_SCI}
\end{figure}

Having established the structural existence of the Pb-dominated cage around Li atoms, we now turn to its dynamic consequences. If Li atoms are indeed confined within such cages, their long-range diffusion must be governed by a series of cage-breaking and escape events, which would manifest as anomalous dynamics. 
The atomic speed and velocity distributions depicted in Fig. \ref{velocity_analysis_SCI} provide the fundamental statistical signal of thermal motion and lay the groundwork for understanding the distinct dynamic roles of Li and Pb within the cage effect framework.
The speed distributions of Li and Pb atoms conform to the Maxwell-Boltzmann distribution, characteristic of a system in thermal equilibrium at 800 K. However, the stark contrast is immediately apparent. The Pb distribution is sharp and narrow, indicative of a low most-probable speed and a confined range of kinetic energies. In contrast, the Li distribution is broad and shallow, signifying a much higher most-probable speed and a significantly wider dispersion of instantaneous kinetic energies.
This discrepancy stems directly from the large difference in atomic mass ($Pb > Li$). At a given temperature, lighter atoms must move faster on average to possess the same average kinetic energy as heavier atoms. This fundamental difference dictates their respective roles in the cage dynamics. Li atoms with inherently higher speeds act as the active high-energy particles incessantly testing and impacting the confines of Pb cages. Pb atoms with slower and more constrained motion, constitute the more static, collective framework of the cage walls.
The velocity component distributions along the x, y, and z directions further corroborate this picture. All three distributions are symmetric about 0 m/s, confirming the system's isotropy and the absence of net directional flow. The Pb velocity components are highly localized, while the Li components are widely dispersed, mirroring the trends observed in the speed distribution.
In summary, the velocity analysis quantifies the innate driving force behind Li atom mobility. The broad distribution of Li velocities ensures that at any given moment, a substantial fraction of Li atoms possess sufficient kinetic energy to initiate a cage-breaking event. This statistical propensity for high-energy excursions is the primary reason why Li, despite being chemically surrounded and confined by Pb, becomes the faster-diffusing species.

\begin{figure}[h]
	\centering
	\subfigure[]{
		\includegraphics[width=1.5in]{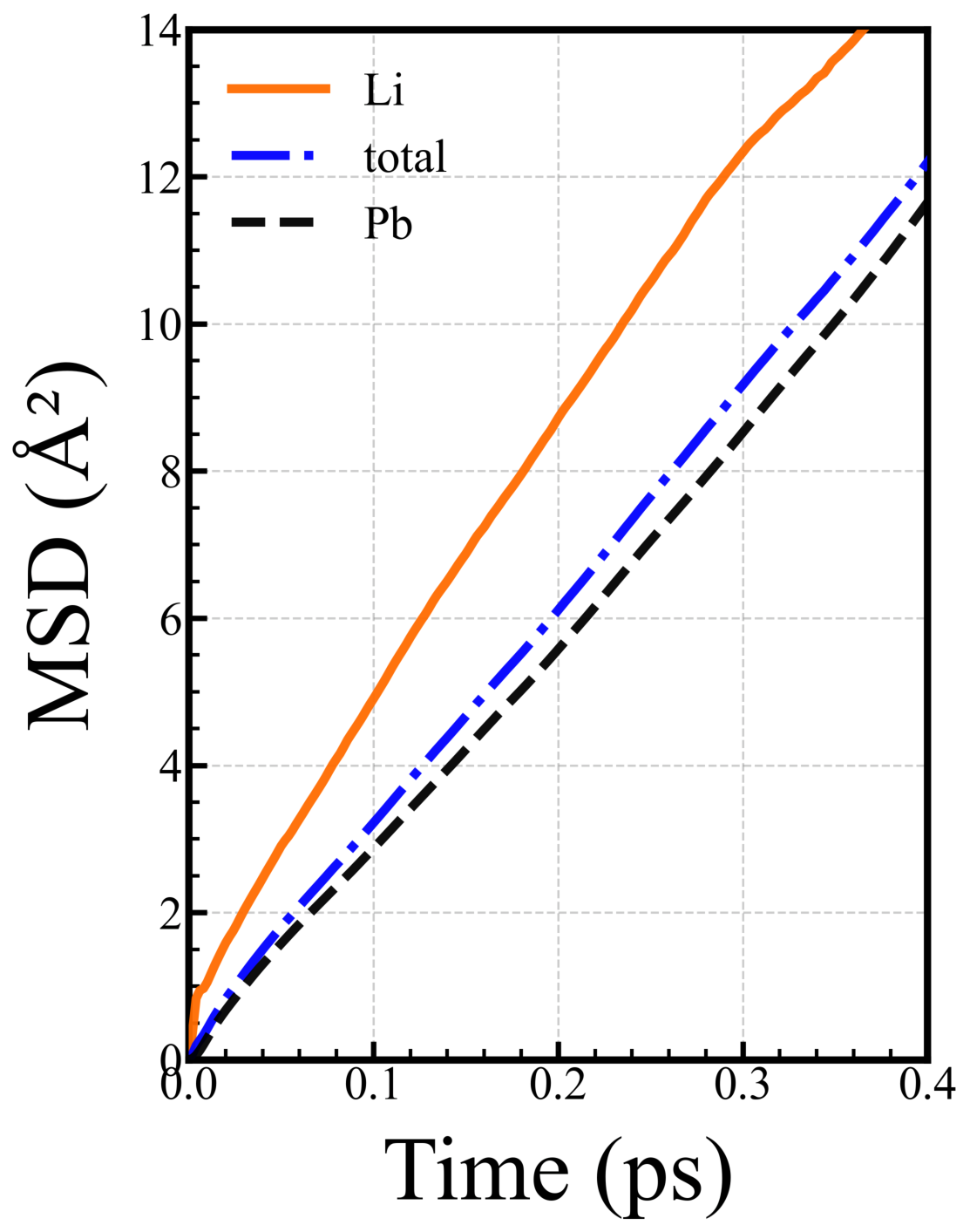}
		\label{msd_alloy_SCI}
	}
	\quad
	\subfigure[]{
		\includegraphics[width=1.5in]{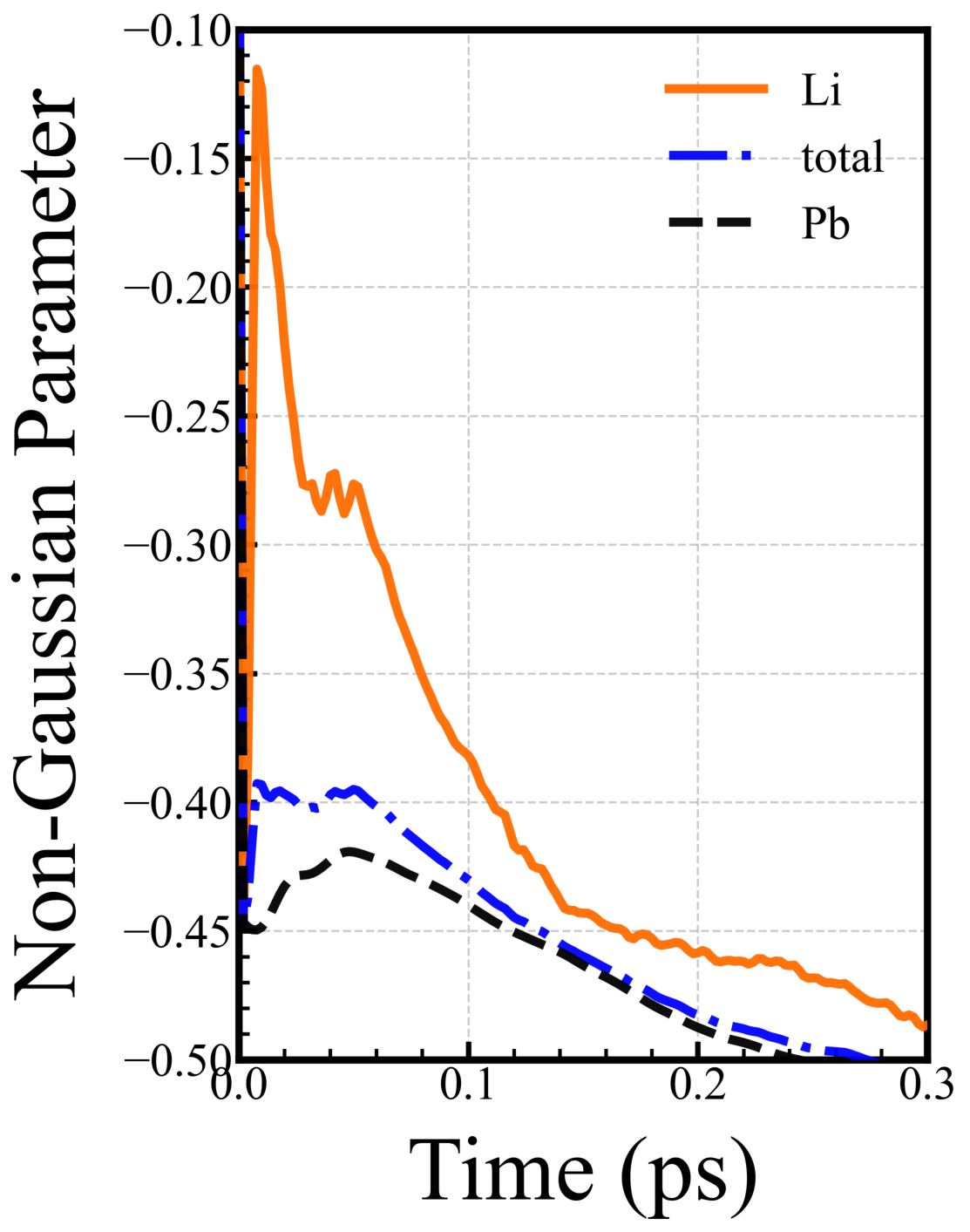}
		\label{ngp_alloy_SCI}
	}
	\caption{(a) Mean speed displacement (MSD) (b) Non-Gaussian parameter.}\label{msd}
\end{figure}

The mean squared displacement (MSD) directly visualizes the starkly contrasting diffusion behaviors of Li and Pb atoms, which is the macroscopic manifestation of the local cage effect. As shown in Fig. \ref{msd_alloy_SCI}, after a short initial ballistic regime ($<$ 0.1 ps), all MSD curves enter a well-defined linear diffusive regime. The slopes of these curves proportional to the diffusion coefficients differ dramatically. The $MS{D_{Li}}$ exhibits the steepest slope, confirming its role as the fastest moving species. In the context of the cage model, this high diffusivity does not imply unhindered motion. Instead, it reflects the successful, intermittent escape of Li atoms from their Pb cages. Each significant upward jump in the $MS{D_{Li}}$ trajectory corresponds to a cage-breaking event, where a Li atom, leveraging its high thermal kinetic energy (as seen in the Fig. \ref{velocity_analysis_SCI}), manages to break its local Li-Pb coordination bonds and relocate to a new cage. In stark contrast, the $MS{D_{Pb}}$ shows the shallowest slope, indicative of sluggish dynamics. This is consistent with Pb atoms primarily forming the static or slowly rearranging network that constitutes the cage walls. Their motion is constrained by strong metallic bonds with neighboring Pb atoms and their large mass further inhibits displacement. The $MS{D_{total}}$ of the system lies between the two, representing the weighted average of these two distinct dynamic modes. The calculated diffusion coefficients quantitatively cement this picture: the Li diffusion coefficient ($1.2 \times {10^{ - 9}} {m^2}/s$) is approximately 3.4 times larger than that of Pb ($3.5 \times {10^{ - 10}}{m^2}/s$). This ratio underscores the effectiveness of the Pb network in constraining its own constituents while still allowing the lighter, more agile Li atoms to perform rapid, jump-like diffusion through its interstices. Therefore, the MSD analysis provides the first direct dynamical evidence that Li diffusion in the 
$Li_{17}Pb_{83}$ alloy is not a simple random walk but a cage-mediated, anomalous process, where the long-range motion of Li is governed by the rate at which it can locally break the cage effect imposed by the Pb matrix. The Non-Gaussian Parameter (NGP) analysis presented in Fig. \ref{ngp_alloy_SCI} moves beyond the average picture provided by the MSD and directly probes the dynamic heterogeneity inherent in the cage-mediated diffusion process. The pronounced temporal evolution and significant differences between the $NG{P_{Li}}$ and $NG{P_{Pb}}$ curves offer compelling evidence for the sporadic, jump-like motion of Li atoms. The most striking feature is the behavior of the $NG{P_{Li}}$. It exhibits large-amplitude oscillations, culminating in a deep negative peak around 0.05 ps. This negative excursion is a classic signature of caged dynamics, a population of Li atoms is temporarily trapped, executing vibrational motions within a confined Pb-cage space, which leads to a sub-diffusive, non-Gaussian displacement distribution. The subsequent strong positive fluctuations in the $NG{P_{Li}}$ are equally telling. They signify the occurrence of rare, large-amplitude displacements precisely the successful cage-breaking events that drive the overall diffusion. Thus, the $NG{P_{Li}}$ trajectory vividly captures the stop-and-go, arrest-and-escape cycle that defines its anomalous diffusion. In stark contrast, the $NG{P_{Pb}}$ remains relatively flat and featureless near zero. This indicates that the motion of Pb atoms is more homogeneous and closer to a Gaussian random-walk diffusion. This is fully consistent with their role as the primary constituents of the slowly rearranging cage matrix. Their movement is a collective, viscous flow rather than the intermittent jumping characteristic of Li. The $NG{P_{total}}$ is dominated by the strong signal from the Li atoms, underscoring that the dynamic heterogeneity of the entire alloy is primarily governed by the cage-breaking and cage-reforming dynamics of the mobile Li species. The NGP analysis provides unambiguous, model-free evidence for the cage effect. It quantifies how the diffusion of Li is not a continuous process but is instead punctuated by periods of temporary imprisonment within the Pb cages, interspersed with rapid, jump-like escapes. This dynamic heterogeneity is the fundamental reason behind the non-Gaussian nature of Li diffusion. Through first-principles calculations, our study clearly reveals this phenomenon in liquid $Li_{17}Pb_{83}$ for the first time, and further attributes the structural origin of this effect to the CSRO of Li-Pb and its specific polar covalent bonding.

\begin{figure}[h]
	\centering
	\subfigure[]{
		\includegraphics[width=1.5in]{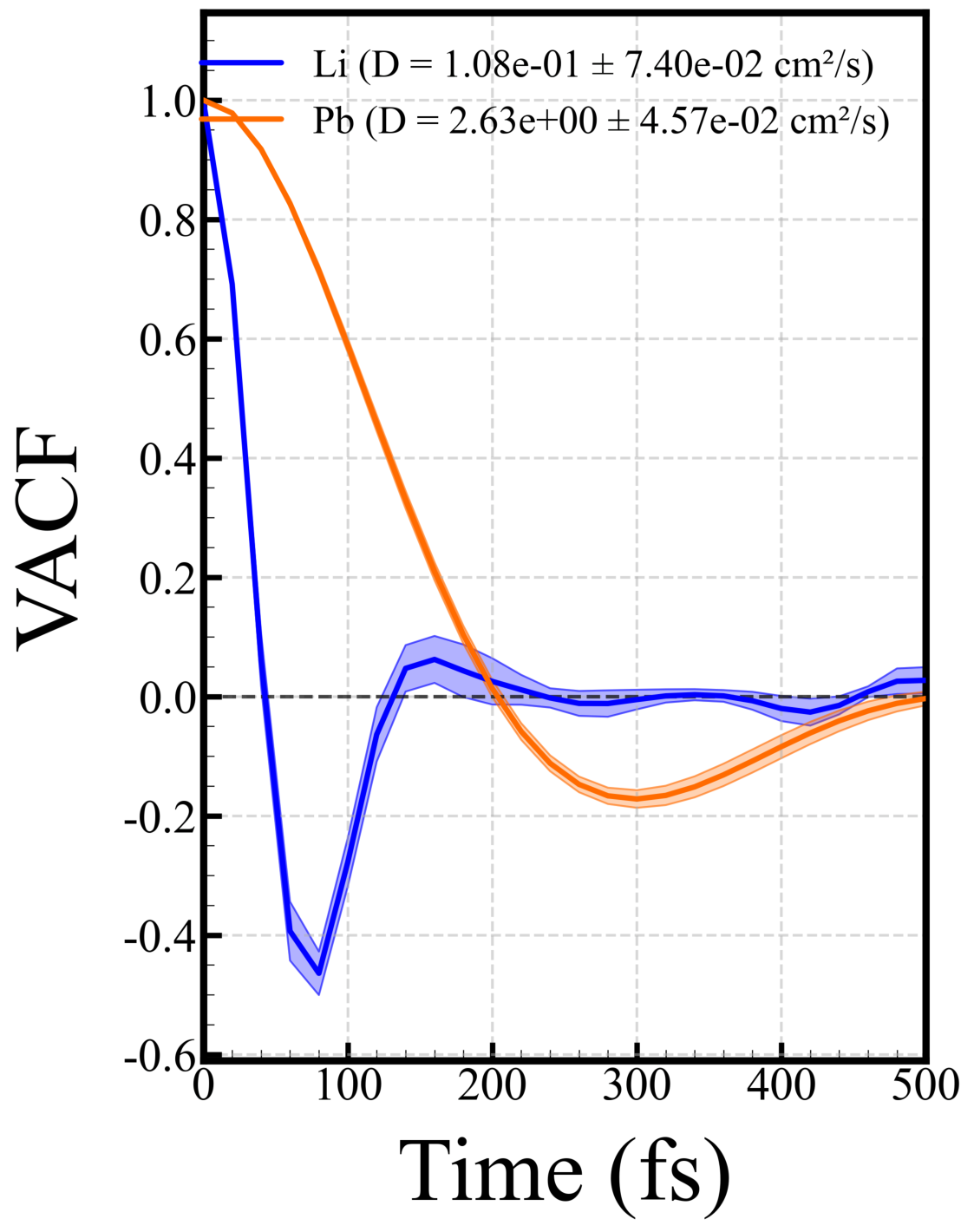}
		\label{combined_vacf_SCI}
	}
	\quad
	\subfigure[]{
		\includegraphics[width=1.5in]{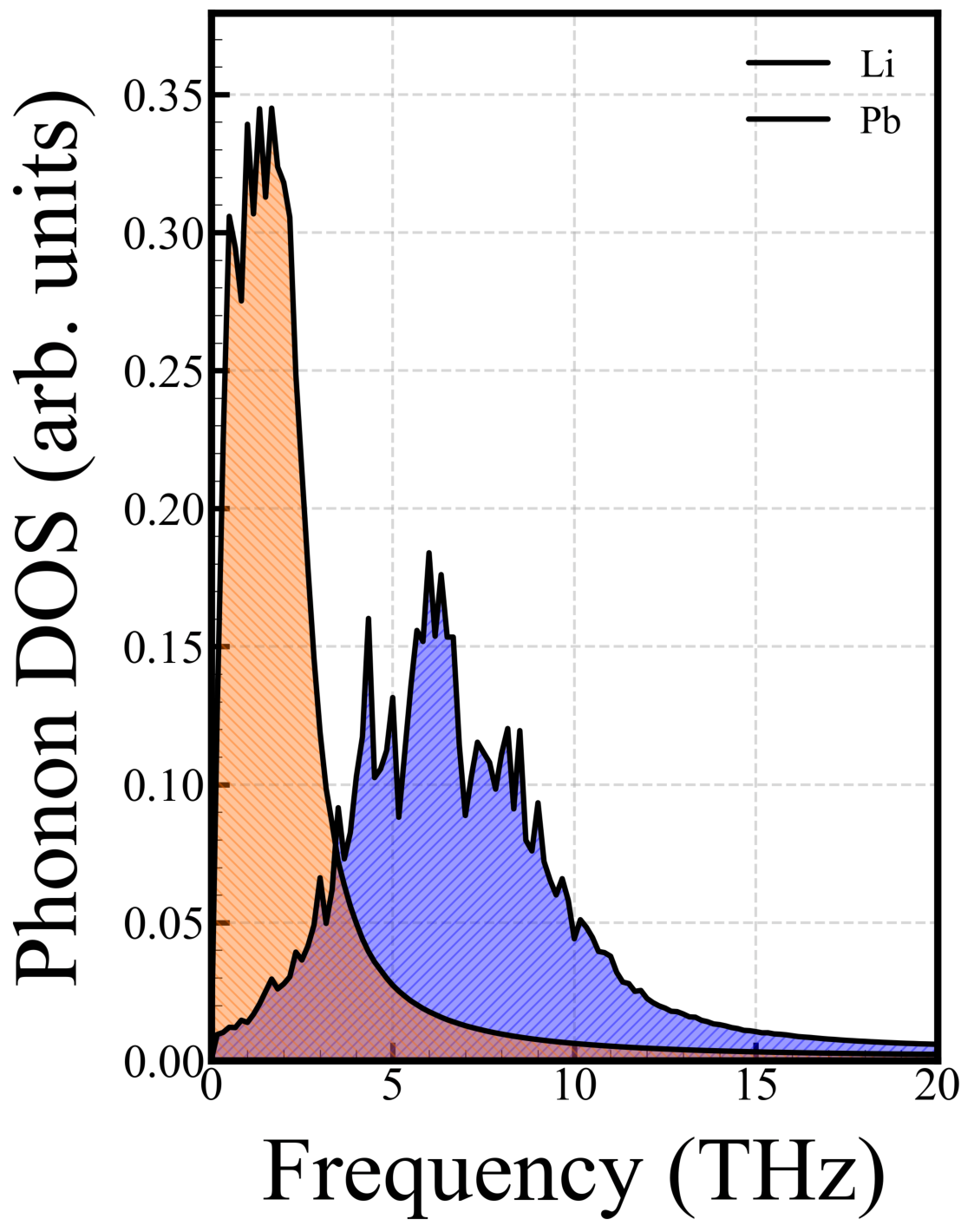}
		\label{combined_phonon_dos_SCI}
	}
	\quad
	\subfigure[]{
		\includegraphics[width=1.5in]{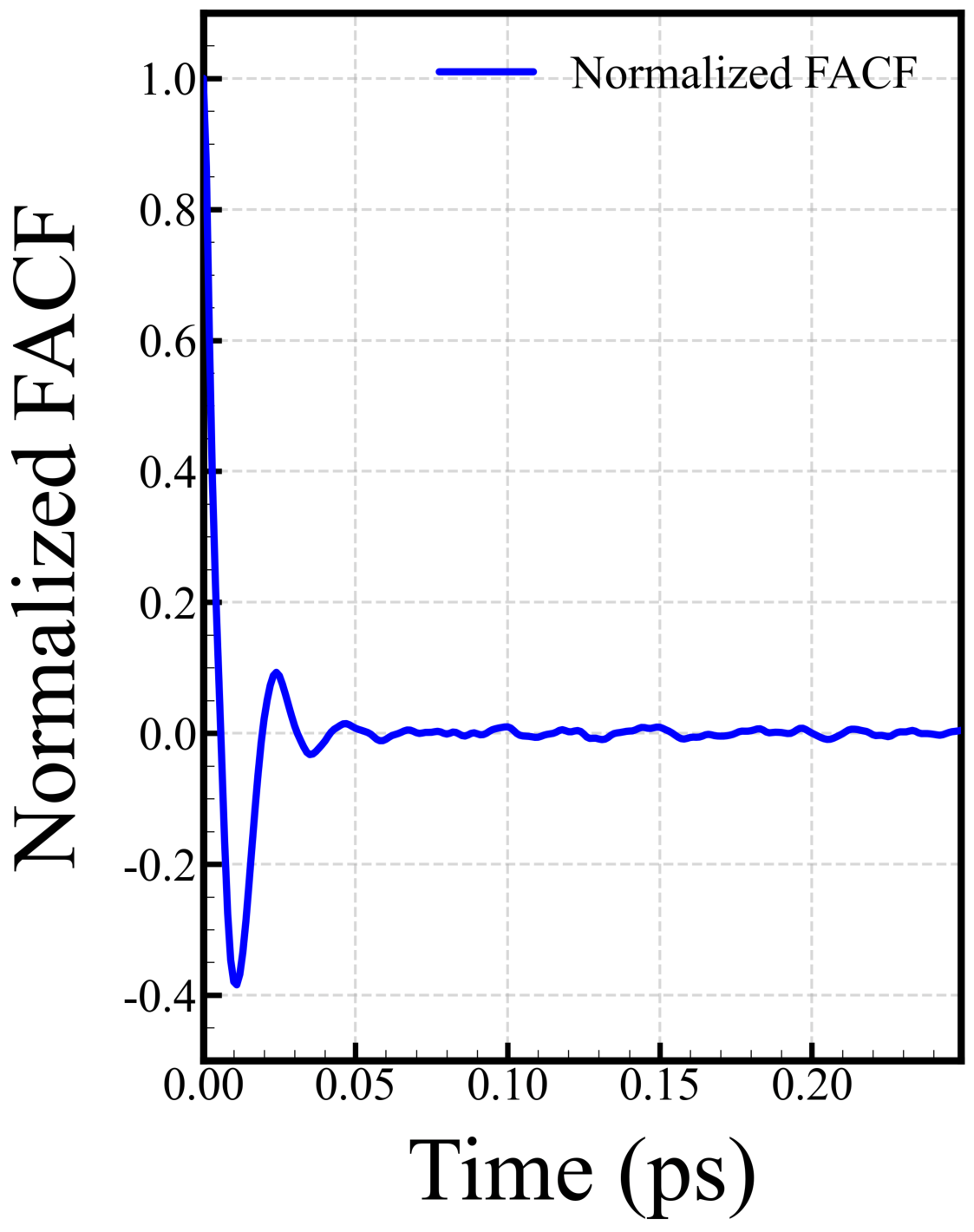}
		\label{facf_SCI}
	}
		\quad
	\subfigure[]{
		\includegraphics[width=1.5in]{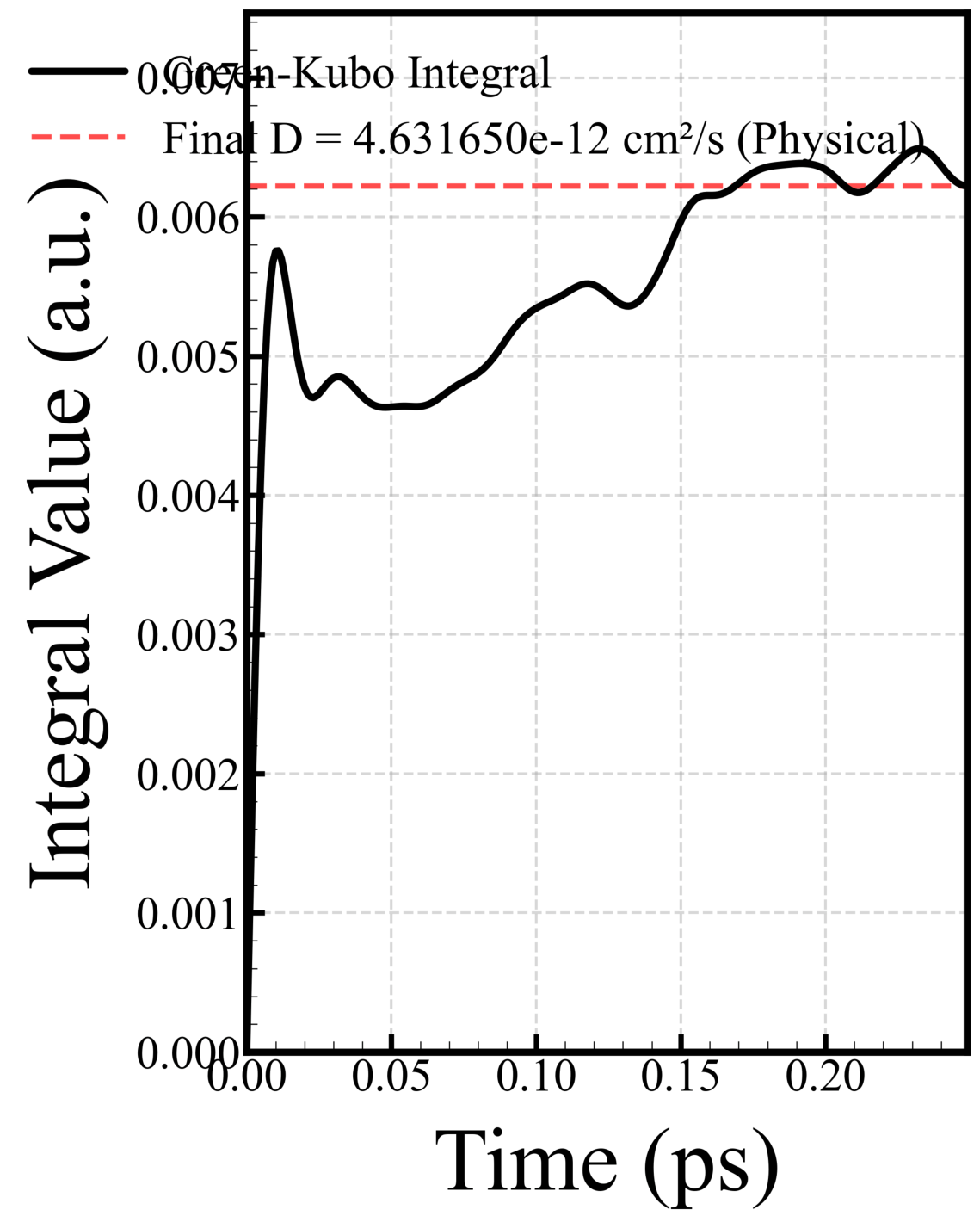}
		\label{diffusion_integral_physical_SCI}
	}
	\quad
	\subfigure[]{
		\includegraphics[width=1.5in]{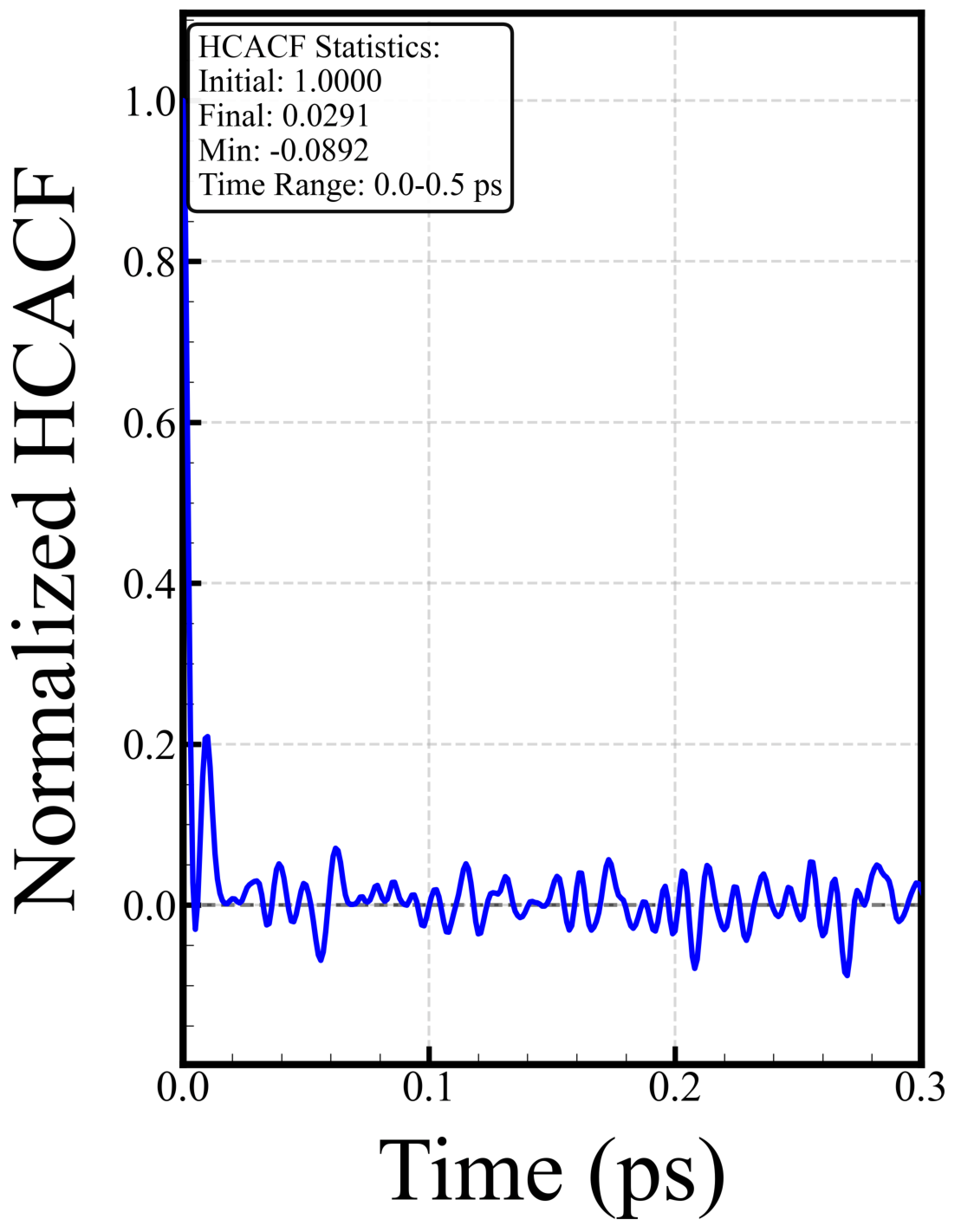}
		\label{hcacf_plot_SCI}
	}
		\quad
	\subfigure[]{
		\includegraphics[width=1.5in]{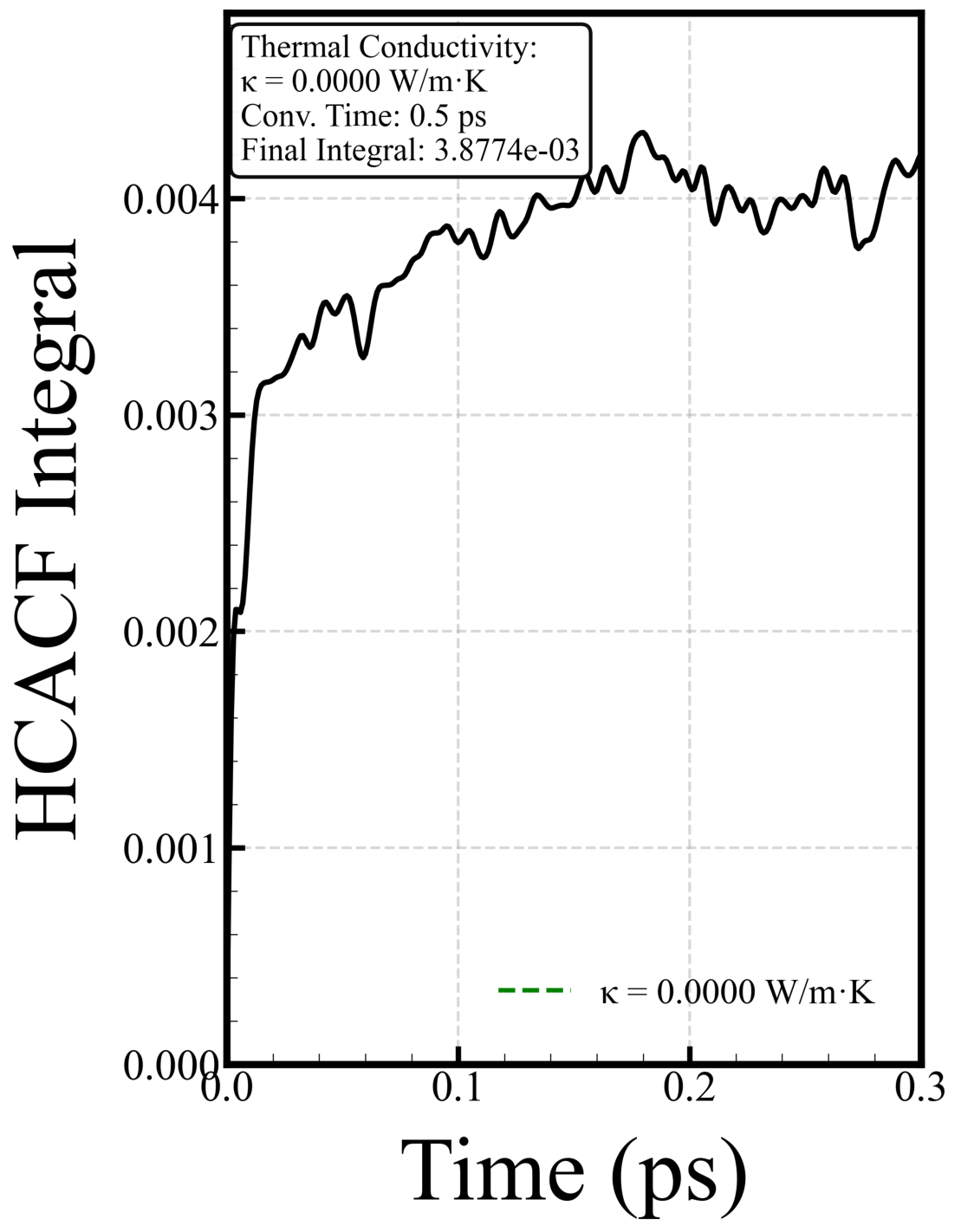}
		\label{thermal_conductivity_convergence_SCI}
	}
	\caption{(a) VACF of Li and Pb (b) PDOS of Li and Pb (c) FACF (d) Integral value of FACF (e) HCACF (f) HCACF Integral.}\label{acf}
\end{figure}

The velocity autocorrelation function (VACF) in Fig. \ref{combined_vacf_SCI} reveals fundamentally distinct dynamic characters. The $VAC{F_{Pb}}$ decays rapidly to zero, consistent with its role as the constituent of a viscous, cage-forming matrix where momentum is quickly randomized by collisions. In stark contrast, the $VAC{F_{Li}}$ exhibits pronounced damped oscillations. This is the classic signature of cage rattling, where a Li atom is temporarily trapped, repeatedly bouncing off the walls of Pb cage before its momentum is randomized upon escape. This oscillatory behavior reflects the vibrational modes within the confinement potential of the cage.
The origin of this difference is further elucidated by the phonon density of states (PDOS) in Fig. \ref{combined_phonon_dos_SCI}. The $PDO{S_{Pb}}$ is highly concentrated in the low-frequency region (< 10 THz), indicative of the collective, low-energy vibrations of the heavy Pb network that forms the cage walls. Conversely, the $PDO{S_{Li}}$ spans a much broader frequency range (0-20 THz), encompassing both the low-frequency cage rattling and higher-frequency local bond-stretching modes. This broad distribution confirms that Li atoms experience a wide spectrum of restoring forces due to the disordered cage environment, which is directly responsible for the complex, oscillatory decay of their VACF.
The force autocorrelation function (FACF) in Fig. \ref{facf_SCI} decays to zero within an exceptionally short timescale (< 0.1 ps). This precipitous drop signifies the extremely short memory of interatomic forces. The physical interpretation within model is clear, the local coordination environment around any atom, the very structure of the cage is undergoing rapid and continuous fluctuations. The force experienced by an atom changes dramatically as its neighboring atoms rearrange, a process that is integral to the cage-breaking and reformation events. The Green-Kubo integral of the FACF in Fig. \ref{diffusion_integral_physical_SCI} converges to a diffusion coefficient consistent with the MSD analysis, validating the dynamical picture derived from these correlation functions.
The heat current autocorrelation function (HCACF) in Fig. \ref{hcacf_plot_SCI} also displays a rapid initial decay, confirming that heat conduction like mass transport, is a diffusive process dominated by short-range interactions. The subsequent small-amplitude oscillatory decay towards zero is highly significant. It suggests the presence of non-random, collective contributions to heat flow that persist for a few picoseconds. It attributes to the dynamical mode separation between the two species. The rapid, jump-like motion of light Li atoms and the localized, low-frequency vibrations of the heavy Pb network act as different heat carriers. Their synergistic yet distinct dynamics introduce oscillatory behavior into the energy correlation. The resulting thermal conductivity in Fig. \ref{thermal_conductivity_convergence_SCI}, calculated from the HCACF integral, is consequently low, as expected for a disordered liquid system where the cage effect strongly scatters heat-carrying phonons and limits the mean free path of diffusive carriers.
In concert, a dynamically disordered system was governed by the local cage effect. The dynamics are dominated by local, short-range events, cage rattling, rapid bond rearrangement and localized energy transfer. The memory of an atom's velocity, the forces acting upon it, and the collective heat flow are all quickly lost due to the persistent structural fluctuations of the liquid cage environment. This multi-angle analysis conclusively shows that the cage is not merely a structural motif but a dynamic entity that fundamentally dictates the momentum, interaction and energy transport properties of the liquid 
$L{i_{17}}P{b_{83}}$ alloy.

\begin{figure}[h]
	\centering
	\includegraphics[width=3.4in]{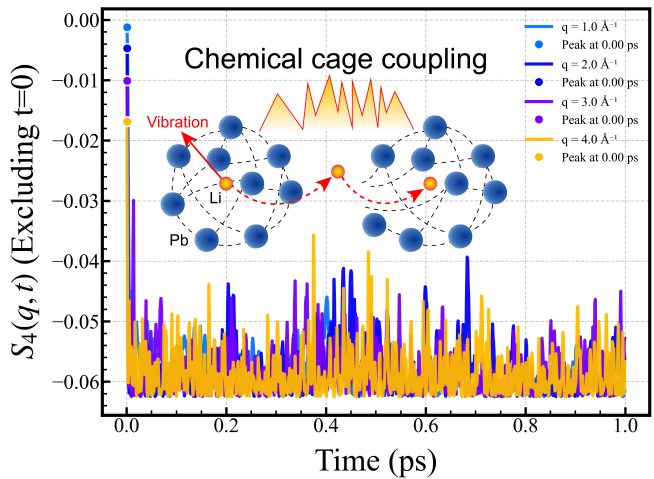}
	\caption{${S_4}\left( {q,t} \right)$ (Four-point dynamic structure factor) analysis of cage coupling effect.}
	\label{S4_detailed_SCI}
\end{figure}

To explore the cooperative dynamics of Li atoms in liquid $Li_{17}Pb_{83}$, we employed the four-point dynamic structure factor ${S_4}\left( {q,t} \right)$ to quantify spatiotemporal correlations between distinct atomic displacements, which directly reveals collective rearrangements and dynamic heterogeneity in liquid $Li_{17}Pb_{83}$. The ${S_4}\left( {q,t} \right)$ is defined to capture correlations between the displacements of distinct atoms (excluding self-correlations), formulated as:
\begin{equation}
	\begin{split}
		S_4(q,t) &= \\
		&\frac{1}{N(N - 1)}\sum_{i \ne j} \left\langle \exp\left[iq \cdot \Delta r_i(t)\right] \exp\left[-iq \cdot \Delta r_i(t)\right] \right\rangle
	\end{split}
\end{equation}
where $\Delta \mathbf{r}_i(t) = \mathbf{r}_i(t) - \mathbf{r}_i(0)$ denotes the displacement of atom $i$ over time $t$, $\mathbf{q}$ is the wavevector  and the angular brackets denote ensemble averaging. A non-negligible $S_4(q,t)$ signal indicates coordinated motion between atoms $i$ and $j$, with the peak position reflecting the characteristic timescale of such cooperative events.
While the NGP highlights the heterogeneous and reveals localized cage-rattling vs. rare escape events, it cannot address whether these cage-breaking events are spatially correlated. This critical gap was resolved by $S_4(q,t)$, which directly probes how the displacement of one Li atom relates to that of its neighbors. 
As shown in Fig. \ref{S4_detailed_SCI}, $S_4(q,t)$ exhibits a pronounced, growing signal that emerges on a timescale matching two key features. The primary relaxation of the ISF marks the breakdown of initial cage constraints, and the positive fluctuations of the NGP signals anomalous diffusion. This temporal alignment provides unambiguous evidence that the rearrangement of one Li-centered cage is not isolated but facilitated by concurrent rearrangements of neighboring cages. Li atoms do not escape their Pb-based chemical cages independently. Instead, their escape is concerted driven by collective reorganization of the surrounding Pb matrix. 
The peak in $S_4(q,t)$ quantifies the characteristic timescale for these cooperative events, underscoring that the chemical cage network is dynamically coupled. This finding directly challenges the single-particle framework of classical MCT, which treats cages as independent entities governed solely by geometric constraints. In contrast, our results reveal a chemically mediated, spatially extended dynamic correlation. Pb’s electronic cage walls stabilized by localized 6s orbitals collectively reorganize to enable Li’s cooperative hopping which adding a critical cage coupling dimension to the physics of liquid alloy dynamics.

\section{Conclusion}
This work systematically investigates the structural, dynamic and bonding characteristics of the liquid $Li_{17}Pb_{83}$ alloy at 800 K. Structurally, the Pb–Pb pairs exhibit the strongest short-range aggregation and local ordering, whereas Li–Li pairs show weaker ordering and greater atomic thermal fluctuations. Dynamically, the MSD and diffusion coefficients reveal that Li atoms diffuse significantly faster ($1.2 \times {10^{ - 9}}$ ${m^2}/s$) than Pb atoms ($3.5 \times {10^{ - 10}}$ ${m^2}/s$), attributable to the lower mass and weaker coordination constraints of Li. The NGP and S4 further indicate strong non-Gaussian behavior and dynamic heterogeneity for Li, while Pb exhibits nearly Gaussian diffusion due to a pronounced cage effect. More importantly, integrating structural analysis and dynamic results, this study proposes for the first time the concept of a chemical bond-directed cooperative cage effect at the atomic scale. This mechanism, well explained by the Mode Coupling Theory, describes Li diffusion as consisting of fast intracage vibrations ($\beta$ relaxation) and slow inter-cage jumps ($\alpha$ relaxation), rather than free diffusion. The double-exponential relaxation in the Li–Pb bond probability further reflects distinct dynamic behaviors between surface and bulk bonds. In summary, this work not only elucidates the anomalous diffusion mechanism in $Li_{17}Pb_{83}$, but also establishes a new theoretical framework for understanding a broad range of systems—from high-temperature alloys to ionic liquids—by incorporating chemical specificity into the core description of liquid dynamics.

\begin{acknowledgments}
This work was supported by the National Natural Science Foundation of China under Grant No. 12305233.
\end{acknowledgments}

\section*{Data Availability Statement}
**************************

\appendix

\nocite{*}
\bibliography{aipsamp}

\end{document}